\documentclass[twocolumn,pra,showpacs,amsmath,amssymb]{revtex4-1}
\usepackage[dvipdfmx]{graphicx}
\usepackage{setspace}
\usepackage{color}
\usepackage{ulem}
\graphicspath{{./figures/}}

\usepackage{amsmath,amsthm,amssymb}
\usepackage{mathrsfs}
\usepackage{braket,comment}
\usepackage{bm}

\usepackage[colorlinks=true, citecolor=blue, urlcolor=blue, linkcolor=blue,
setpagesize=false, bookmarks=false
]{hyperref}

\begin{document}
\title{Thermally enhanced Majorana-mediated spin transport in the Kitaev model}

\author{Hirokazu Taguchi}
\affiliation{Department of Physics, Tokyo Institute of Technology, Meguro, Tokyo 152-8551, Japan}
\author{Yuta Murakami}
\affiliation{Department of Physics, Tokyo Institute of Technology, Meguro, Tokyo 152-8551, Japan}
\author{Akihisa Koga}
\affiliation{Department of Physics, Tokyo Institute of Technology, Meguro, Tokyo 152-8551, Japan}
\date{\today}

\begin{abstract}
  We study how stable the Majorana-mediated spin transport
  in a quantum spin Kitaev model is against thermal fluctuations.
  Using the time-dependent thermal pure quantum state method, we examine
  finite-temperature spin dynamics in the Kitaev model.
  The model exhibits two characteristic temperatures $T_L$ and $T_H$, which correspond to energy scales of the local flux and the itinerant Majorana fermion, respectively.
  At low temperatures $(T\ll T_L)$, an almost flux-free state is realized and
  the spin excitation propagates in a similar way to that for the ground state.
  Namely, after the magnetic pulse is introduced at one of the edges,
  the itinerant Majorana fermions propagate the spin excitations
  even through the quantum spin liquid state region,
  and oscillations in the spin moment appear in the other edge with a tiny magnetic field.
  When $T\sim T_L$, larger oscillations in the spin moments are induced
  in the other edge, compared to the results at the ground state.
  At higher temperatures,
  excited $Z_2$ fluxes disturb the coherent motion of
  the itinerant Majorana fermions,
  which suppresses the spin propagation.
  Our results demonstrate a crucial role of thermal fluctuations
  in the Majorana-mediated spin transport.
\end{abstract}

\maketitle

\section{Introduction}

Recently, spin transport has been attracting much interest.
One of the examples is the spin current induced
by a polarized electric current in the ferromagnetic metals
~\cite{Slonczewski_1989, Berger_1996, Slonczewski_1996, Bhat_2000, Tsoi2000, Konig_2001, Spintronics_2004, Ogawa2016}.
Another example is the spin current in the magnetic insulators,
where magnons carry spins without the electric current
~\cite{Tsui_1971, Moodera_1995, Kajiwara_2010, Cornelissen_2015}.
In both cases, the spin current flows in materials with magnetic orders.
On the other hand, it has been revealed that the spin transport
is also realized in quantum spin liquids (QSLs)
~\cite{Minakawa_2020,KogaSpin_2020,Taguchi_2021},
where no magnetic order is realized due to strong quantum fluctuations
~\cite{ANDERSON1973153, Read_1989, Wen_1991, qsl1, Chen_2013, qsl2, Zhou_2017}.
One of the typical examples is provided by
an antiferromagnetic $S = 1/2$ Heisenberg chain.
The anisotropic negative spin Seebeck effect in a candidate material $\text{Sr}_2 \text{CuO}_3$ indicates the spin current mediated
by spinons~\cite{SpinonSpinCurrent},
which are magnetic elementary excitations in this system.

Another interesting playground for QSLs is given by the Kitaev model~\cite{Kitaev_model},
which is composed of direction-dependent Ising interaction between $S=1/2$ spins
on the honeycomb lattice.
In the model, quantum spins are fractionalized into
itinerant Majorana fermions and local fluxes due to quantum many-body effects.
The itinerant Majorana fermions have been observed as a half quantized plateau
in the thermal quantum Hall experiments~\cite{Nasu_2017, Kasahara_2018} in a candidate $\alpha$-RuCl$_3$~\cite{Plumb_2014}.
Furthermore, it has been reported that the itinerant Majorana fermions
play a crucial role for the spin transport without spin oscillations
~\cite{Minakawa_2020,KogaSpin_2020,Taguchi_2021}.
It is known that Majorana and flux excitations have distinct energy scales,
which leads to interesting thermodynamic properties such as
the double peaks in the specific heat and
the plateau in the entropy~\cite{Nasu_2014,Nasu_2015}.
Therefore it is highly desired to clarify how stable such Majorana related phenomena are
against thermal fluctuations.
This should be important to realize spintronics devices with Majorana fermions.

To answer this question, we deal with the Kitaev model with edges and consider the spin transport at finite temperatures.
By means of the time-dependent thermal pure quantum (TPQ) state method
~\cite{Sugiura_2012, Sugiura_2013, Endo_2018},
we examine the dynamics of the system after
the magnetic pulse is introduced at one of the edges.
Then, we discuss how thermal fluctuations affect the Majorana-mediated spin transport.

The paper is organized as follows.
In Sec.~\ref{sec:model}, we introduce the Kitaev model on the honeycomb lattice and
explain the time-dependent TPQ method.
In Sec.~\ref{sec:result}, we discuss how stable the spin propagation in the Kitaev model is against thermal fluctuations.
A summary is given in the last section.

\section{Model and Method} \label{sec:model}
\begin{figure}[htb]
  \centering
  \includegraphics[width=\linewidth]{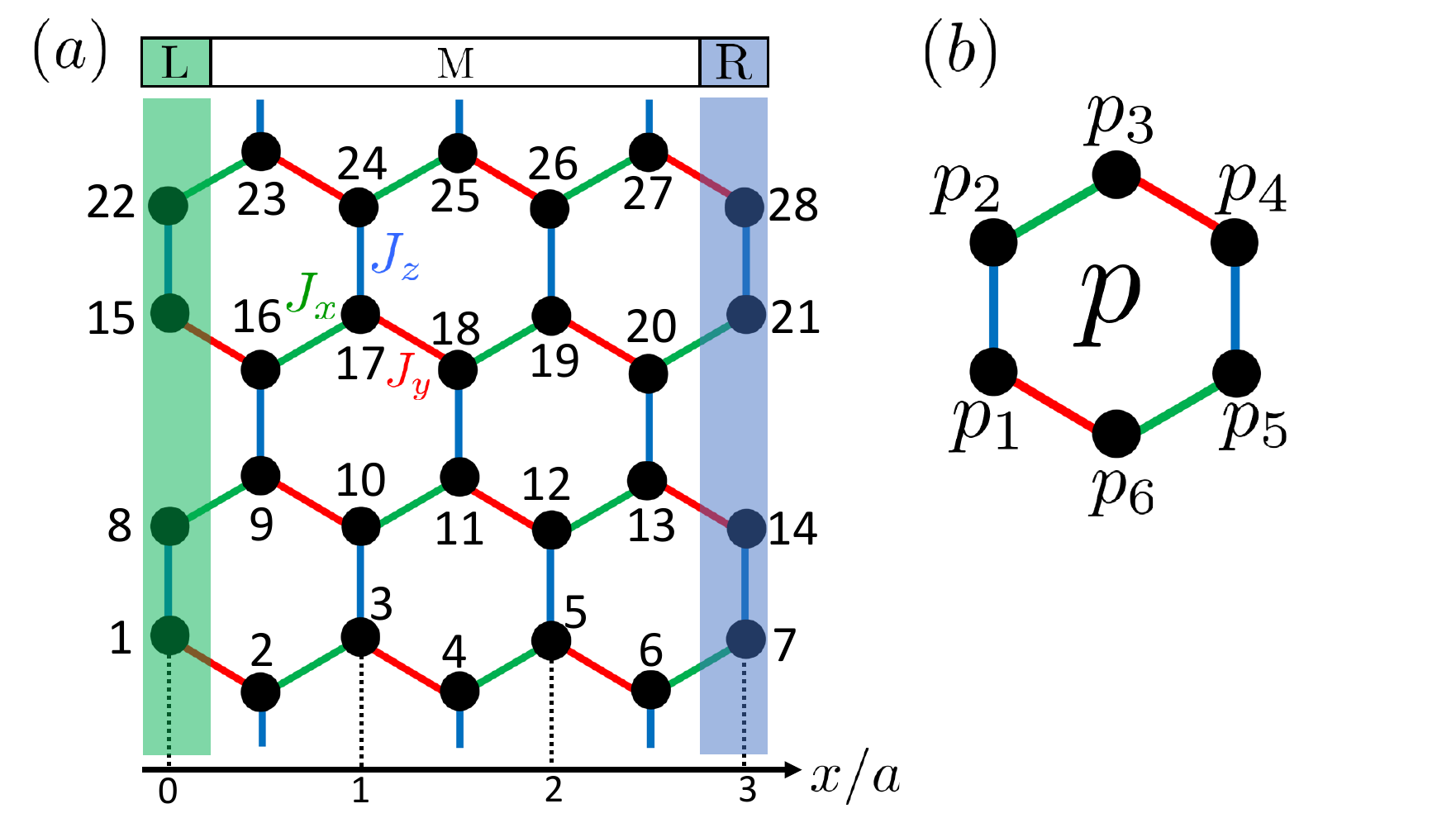}
  \caption{
    (a) Kitaev model with armchair edges.
    Green, red, and blue lines indicate $x$-, $y$-, and $z$-bonds, respectively.
    Its bond length is $a/\sqrt{3}$.
    The static magnetic field $h_R$ is applied in the right $(R)$ region, and
    no magnetic field is applied in the middle (M) region.
    Time-dependent pulsed magnetic field is introduced in the left (L) region.
    (b) Plaquette $p$ with sites marked $p_1, p_2, \cdots,p_6$ shown for
    the operator $W_p$.
  }
  \label{fig:system}
\end{figure}

We consider the Kitaev model
on a two-dimensional honeycomb lattice,
which is given by the following Hamiltonian as
\begin{align}
  &H_K = -J\sum_{\braket{i,j}_x}S_i^x S_j^x-J\sum_{\braket{i,j}_y}S_i^y S_j^y-J\sum_{\braket{i,j}_z}S_i^z S_j^z,
\end{align}
where $\braket{i, j}_\mu$ indicates the nearest-neighbor pair on the $\mu(=x,y,z)$-bonds.
The $x$-, $y$-, and $z$-bonds are shown as green, red, and blue lines
in Fig.~\ref{fig:system}.
$S_i^\mu$ is the $\mu$ component of an $S = 1/2$ spin operator at the $i$th site and
$J$ is the exchange coupling between the nearest-neighbor spins.

An important feature is that the Kitaev model has the local conserved quantities.
The operator $W_p$ is defined as
\begin{eqnarray}
  W_p=2^6\cdot S_{p_1}^x S_{p_2}^y S_{p_3}^z S_{p_4}^x S_{p_5}^y S_{p_6}^z,
\end{eqnarray}
where $p_i\;(i=1, 2, \cdots, 6)$ is the site in the plaquette $p$ [see Fig.~\ref{fig:system}(b)].
Since $[H_K, W_p]=0, [W_p, W_q]=0$, and $W_p^2=1$,
the operator $W_p$ is a $Z_2$ local conserved quantity.
Each eigenstate of the Kitaev model is classified by
the Hilbert space specified by a set $\{w_p\}$,
where $w_p(=\pm 1)$ is the eigenvalue of $W_p$.
Since the spin operator changes the sign of the corresponding eigenvalues $w_p$
for a certain state,
the existence of the local conserved quantity guarantees the absence
of local magnetic moments $\langle S^\mu_i\rangle$ and
long-range spin-spin correlations $\langle C_{ij}^\mu \rangle$ in the Kitaev model,
where $C_{ij}^\mu=S_i^\mu S_j^\mu$.
The ground state is realized in the space with $w_p=1$ for each plaquette,
which can be regarded as the flux-free space~\cite{Kitaev_model}.
Besides the flux degrees of freedom, the other remains: itinerant Majorana fermions.
It is known that the gapless dispersion with the velocity $v[=(\sqrt{3}/4)aJ]$
appears in the itinerant Majorana excitation in the flux-free space, where $a$ is a lattice constant.
It is also known that the finite energy is needed to create
adjacent fluxes in the system~\cite{Kitaev_model}.
The energy scales of the itinerant Majorana and the flux excitations
are distinct from each other. In the followings, we discuss how the energy difference
affects the spin transport at finite temperatures.

To study the spin transport in the Kitaev model,
we treat the system with armchair edges, as
shown in Fig.~\ref{fig:system}$(a)$.
The system is composed of L, M, and R regions,
where the distinct magnetic fields are applied in the $z$ direction.
In the L region on the left edge, a time-dependent pulsed magnetic field $h_L(t)$
is introduced around $t=0$.
No magnetic field is applied to the M region, while the static magnetic field $h_R$
is applied to the R region.
The model Hamiltonian is given as
\begin{align}
  &H(t) = H_0 + H_1(t), \label{eq:model} \\
  &H_0 = H_K-h_R\sum_{i \in R}S_i^z,  \\
  &H_1(t) = - h_L(t)\sum_{i \in L}S_i^z.
\end{align}
We note that in the regions under the finite magnetic field,
the local operator $W_p$ is no longer a conserved quantity.
For example, the local operator on the plaquette composed of the sites
(5, 12, 13, 14, 7, 6) shown in Fig.~\ref{fig:system}(a)
does not commute with $-h_RS_{7}^z$.
Therefore, in general, this leads to the finite magnetizations in the R region.

In the paper, we examine the real-time dynamics in the model
at finite temperatures
after the magnetic pulse is introduced in the L region.
The expectation value at time $t$ for an operator $\hat{O}$ is given as
\begin{align}
  \braket{\hat{O}(t)} = \frac{1}{Z_0}\text{Tr}\left[ \hat{O}(t) e^{- \beta H_0} \right], \label{eq:B1}
\end{align}
where $\beta=1/T$, $T$ is the temperature,
$Z_0(=\text{Tr}\left[ e^{- \beta H_0} \right])$ is the partition function
and $\hat{O}(t) = U^\dag(t)\hat{O}U(t)$ with the time-evolution operator $U(t)$.
At zero temperature ($T=0$),
the localized $Z_2$ fluxes freeze into the flux-free state,
and the Majorana mean-field approach should work to evaluate the expectation values
~\cite{Nasu_2018, Liang_2018, Nasu_2019,  Minakawa_2020, Taguchi_2021}.
On the other hand, at finite temperatures, the above method should be hard
to treat both excitations with distinct energy scales.
Thus, we use the TPQ state method ~\cite{Sugiura_2012, Sugiura_2013},
where local quantities are efficiently evaluated
without the trace calculations
~\cite{Wietek_2019, Hickey_2020, Wietek_2021, Shackleton_2021, Hickey_2021, Sala2021, Suzuki2021}.
An important point is that this numerical method
takes several energy scales into account on equal footing,
and thereby has been successfully used
in several systems such as the Heisenberg model on frustrated lattices
~\cite{Sugiura_2012, Sugiura_2013, Yamaji_2016, Endo_2018, Suzuki_2019, Schaefer, Shimokawa}
and the Kitaev models
~\cite{Tomishige_2018, Nakauchi_2018, KogaS1_2018, Oitmaa_2018, KogaMix_2019, Hickey_2019, Morita_2020}.

In the TPQ method, the expectation value (\ref{eq:B1}) is described
by means of the TPQ state as
\begin{eqnarray}
  \braket{\hat{O}(t)} &=& \langle\Psi_T| \hat{O}(t)|\Psi_T\rangle, \notag \\
  &=&\langle \Psi_T(t)| \hat{O} |\Psi_T(t)\rangle,
\end{eqnarray}
where $|\Psi_T\rangle$ is the TPQ state at the temperature $T$
and $|\Psi_T(t)\rangle=U(t)|\Psi_T\rangle$.
The time-evolution of the physical quantities can be evaluated
by the time-evolution of the TPQ state~\cite{Endo_2018}.

Here, we briefly explain the TPQ method.
A TPQ state at $T\rightarrow\infty$ is simply given by a random vector,
\begin{eqnarray}
  |\Psi_0\rangle = \sum c_i |i\rangle,~\label{random}
  \end{eqnarray}
where $\{ c_i\}$ is a set of random complex numbers satisfying $\sum_i |c_i|^2=1$
and $|i\rangle$ is an arbitrary Hilbert basis.
By multiplying a certain TPQ state by the Hamiltonian,
the TPQ states at lower temperatures are constructed.
The $k$th TPQ state is represented as
\begin{eqnarray}
  |\Psi_k\rangle = \frac{(L-H_0)|\Psi_{k-1}\rangle}{||(L-H_0)|\Psi_{k-1}\rangle||},
\end{eqnarray}
where $L$ is a constant value, which is larger than the maximum
eigenvalue of the Hamiltonian $H_0$.
The corresponding temperature is given by
\begin{eqnarray}
  T_k=\frac{L-E_k}{2k},
\end{eqnarray}
where $E_k(=\langle\Psi_k|H_0|\Psi_k\rangle)$ is the internal energy.
The thermodynamic quantities such as entropy and specific heat
can be obtained from the internal energy and temperature.

We repeat this procedure until $T_k=T$ and
obtain the TPQ state $|\Psi_T\rangle$.
Then, we calculate the time-evolution of the TPQ state $\ket{\Psi_T(t)}$ 
in terms of
the Lanczos time-evolution methods
~\cite{Park_1986, Saad_1992, Druskin_1995, Marlis_1997, Hochbruck_1998, Hochbruck1999}.
We can efficiently obtain the expectation value $\langle \hat{O}(t)\rangle$.
When we discuss the real-time dynamics by applying the pulsed magnetic field,
it is useful to consider a change in the quantities as,
\begin{eqnarray}
  \Delta O(t)=\langle \hat{O}(t)\rangle-\langle \hat{O} \rangle_0,\label{Delta}
\end{eqnarray}
where $\langle \cdots \rangle_0$ is the expectation value
for the static Hamiltonian $H_0$.

When the TPQ method is applied to the finite cluster,
the obtained results are sensitive to its size and/or shape.
This is due to, at least, two effects.
One of them is that
low energy properties in the thermodynamic limit cannot be described correctly
in terms of finite clusters.
Therefore, the large system size dependence of the physical quantities appears at low temperatures
although the TPQ method reproduces the correct results at higher temperatures.
The other is the random dependence in the initial TPQ state.
This should become negligible, by taking a statistical average of
the results for independent TPQ states.
Nevertheless, we sometimes meet with difficulty in evaluating time-dependent quantities,
since each TPQ state is not an eigenstate of the Hamiltonian.
Namely, ill oscillations appear in the physical quantities with respect to time even without time-dependent perturbations,
unless the quantities are conserved ones.
Although this oscillation should be neglected in the statistical average,
the sample dependence is somewhat large even at high temperatures.
To avoid this problem, we construct two time-dependent TPQ states from
the common TPQ state as, $|\Psi_T(t)\rangle$ and $|\Psi_T^0(t)\rangle=U_0(t)|\Psi_T\rangle$,
where $U_0(t)$ is the time-evolution operator for the system described by $H_0$.
Then, we calculate $\langle \hat{O}(t)\rangle_0=\langle \Psi_T^0(t)|\hat{O}|\Psi_T^0(t)\rangle$ instead of $\langle \hat{O} \rangle_0$
and evaluate the change in the quantities (\ref{Delta}),
where unphysical oscillations should be cancelled.
This allows us to obtain $\Delta O(t)$ efficiently and
to discuss correctly how the external field affects the Kitaev system
at finite temperatures.
We have confirmed that, in the 16-site cluster,
our TPQ results are in good agreements with the results obtained
by the finite-temperature exact diagonalization (not shown).

\begin{figure}[t]
  \centering
  \includegraphics[width=\linewidth]{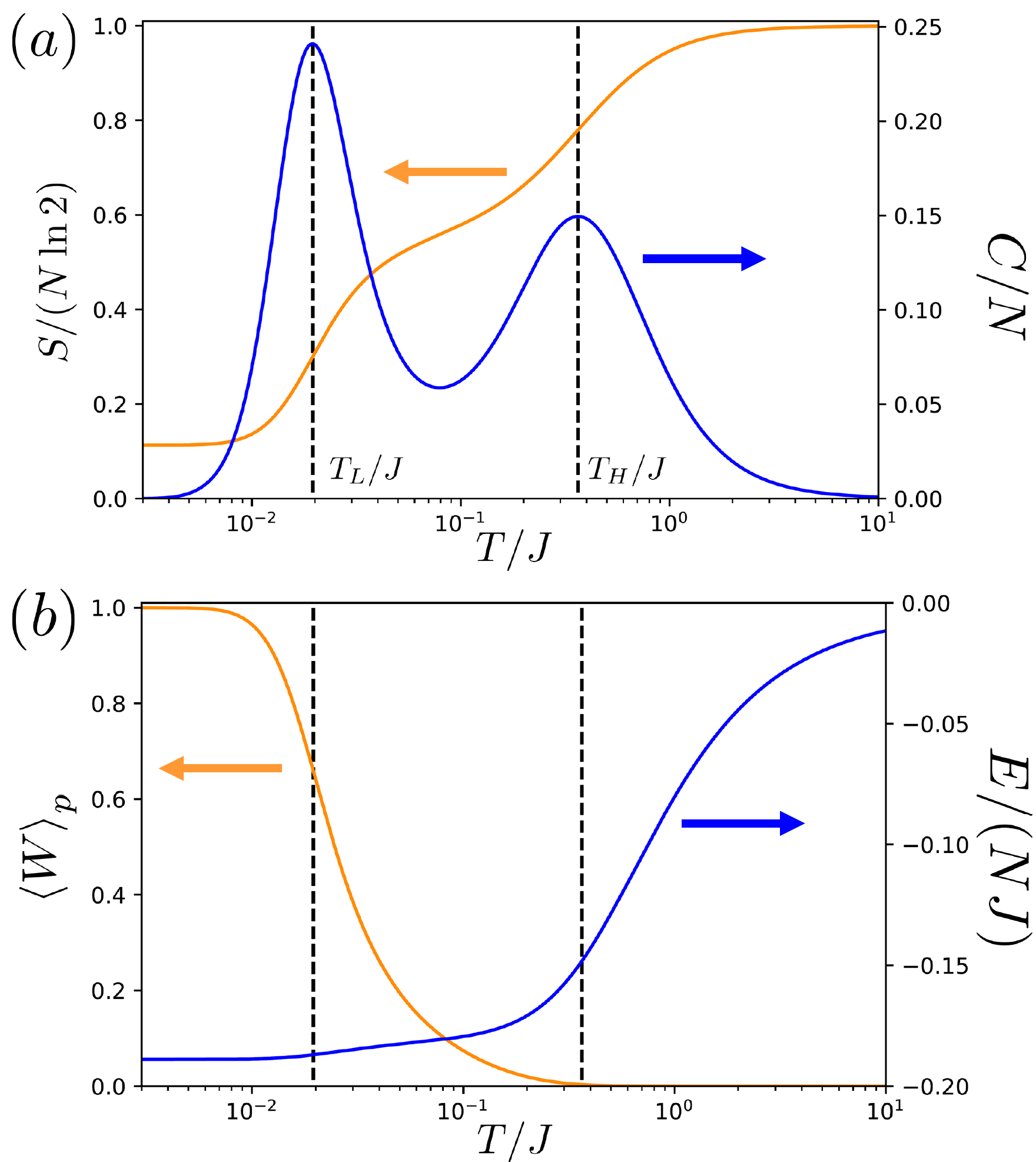}
  \caption{(a) Specific heat $C$ (blue line), entropy $S$ (orange line),
    and (b) internal energy $E$ (blue line) and
    expectation values of local conserved quantities $\braket{W}_p$ (orange line)
    as a function of $T/J$ in the Kitaev system ($N=28$) with the armchair edges described by $H_0$.
  }
  \label{fig:HeatWpN24}
\end{figure}
In the present study, we mainly consider the Kitaev model with $N=28$,
where $N$ is the total number of sites.
The model is schematically shown in Fig.~\ref{fig:system}(a).
The static magnetic field in the R region is set to be $h_R=0.01J$,
which is smaller than the critical values $h_c$
~\cite{Nasu_2018,Liang_2018, Ido_2020}.
Before discussing the time evolution, we first demonstrate equilibrium quantities of the Kitaev model.
Applying the TPQ method to the Hamiltonian $H_0$ with $h_R/J=0.01$
on the 28-site cluster with armchair edges,
we obtain the entropy $S$, specific heat $C$, internal energy $E$,
and expectation value $\langle W_p\rangle$.
The results are shown in Fig.~\ref{fig:HeatWpN24}.
We clearly find double peaks in the specific heat
at $T_L/J\sim 0.019$ and $T_H/J\sim 0.36$,
and shoulder behavior in the entropy around $T/J\sim 0.1$.
It is also found that $E$ is largely changed around $T_H$,
while $\langle W_p\rangle$ is changed around $T_L$.
These results are consistent with the fact that
$T_L$ and $T_H$ correspond to typical energy scales of the local fluxes and the itinerant Majorana fermions, respectively.
The residual entropy originates from the existence of the edge states
in the cluster.
As mentioned above, low temperature properties
are sensitive to the cluster since
low-energy excitations depend on the size and/or shape.
Nevertheless, the spin fractionalization inherent in the Kitaev model
can be captured even in the 28-site system with edges.
This allows us to discuss how thermal fluctuations affect
the spin propagation in the Kitaev model qualitatively.

\section{Results}\label{sec:result}

Now, we study the real-time dynamics of the Kitaev system at finite temperatures
after the Gaussian magnetic pulse is introduced in the L region.
The form of the Gaussian pulse is given as
\begin{align}
  h_L(t) = \frac{A}{\sqrt{2\pi}\sigma}\exp{\left[\frac{t^2}{2\sigma^2} \right]},
  \label{eq:hLt}
\end{align}
where $A$ and $\sigma$ are strength and width of the pulse.
Here, we set $\sigma = 2/J$ and $A = 1$.
Taking average over more than hundred independent TPQ states,
we calculate the time-evolution of local physical quantities.
It is known that the spin transport through the Kitaev QSL region is mediated
by the Majorana fermions~\cite{Minakawa_2020, Taguchi_2021}.
To avoid discussions for the reflection around the right edge,
we define the arrival time of oscillations triggered by the magnetic pulse at $x$ as $t^*=x/v$,
where $x$ is the coordinate of the $i$th site or the midpoint of the bond [see Fig.~\ref{fig:system}(a)].

\begin{figure}[t]
  \centering
  \includegraphics[width=\linewidth]{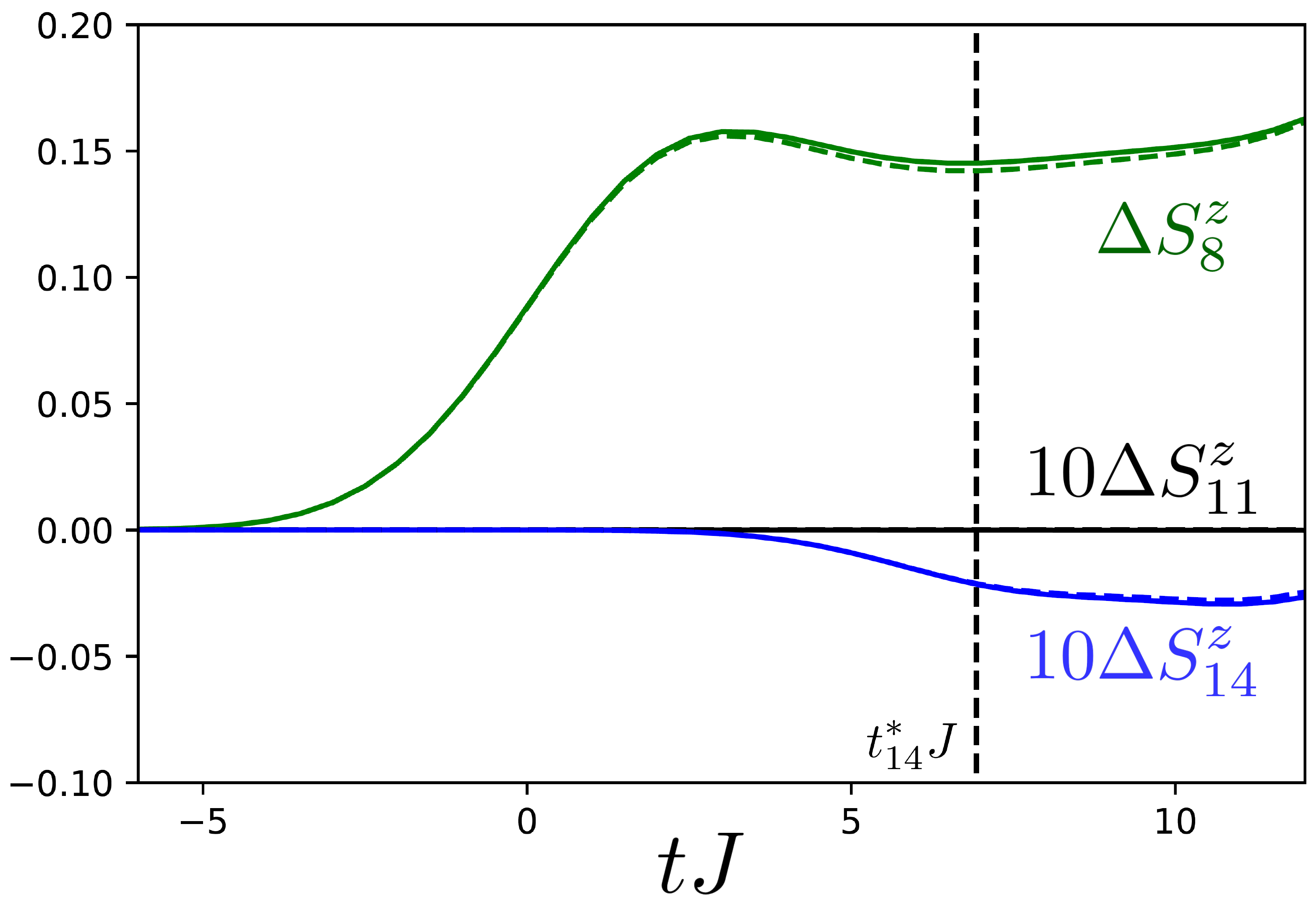}
  \caption{Real-time evolution of the changes in the local magnetizations
    $\Delta S^z_{i}$ in the system with $h_R/J=0.01$ after
    the introduction of the pulsed magnetic field
    with $A=1$ and $\sigma=2/J$.
    Solid and dashed lines represent the results at the temperatures
    $T/J = 0.01$ and $T/J=0$, respectively.
    For clarity, $\Delta S^z_{i}(t)$ for the sites 11 and 14 are plotted
    on a scale of 10 times.
  }
  \label{fig:lowT}
\end{figure}

Figure~\ref{fig:lowT} shows the change in the local magnetizations
in the L, M, and R regions at $T/J=0.01$.
In the L region (site 8),
no magnetic field is applied at $t\rightarrow -\infty$,
and thus no magnetic moment appears.
We find that the pulsed magnetic field induces the magnetic moment $\Delta S^z_8(t)$
at the same time as the pulse is introduced.
On the other hand, no magnetic moments are induced in the M region (site 11).
This is consistent with the fact that the existence of local conserved quantities
guarantees the absence of the magnetic moments even after the magnetic pulse is introduced.
In the R region, the tiny static magnetic field $h_R$ is applied and
the magnetic moment appears with $\langle S^z_{14}\rangle \sim 0.092$ at $t=-\infty$.
We find that
the spin oscillation is induced at the site 14 around $t\sim 3/J$.
This means that the wave packet triggered by the magnetic pulse in the L region reaches
the R region through the M region without spin oscillations.
The peculiar spin transport is mediated by itinerant Majorana fermions~\cite{Minakawa_2020}.
We also apply the exact diagonalization to this system and
calculate the spin oscillation at zero temperature.
The obtained results are shown as the dashed lines in Fig.~\ref{fig:lowT}.
We find that the spin oscillation for the ground state is
slightly different from that at $T/J=0.01$.
This should imply that few excited fluxes influence the motion of the itinerant Majorana fermions.

\begin{figure}[t]
  \centering
  \includegraphics[width=\linewidth]{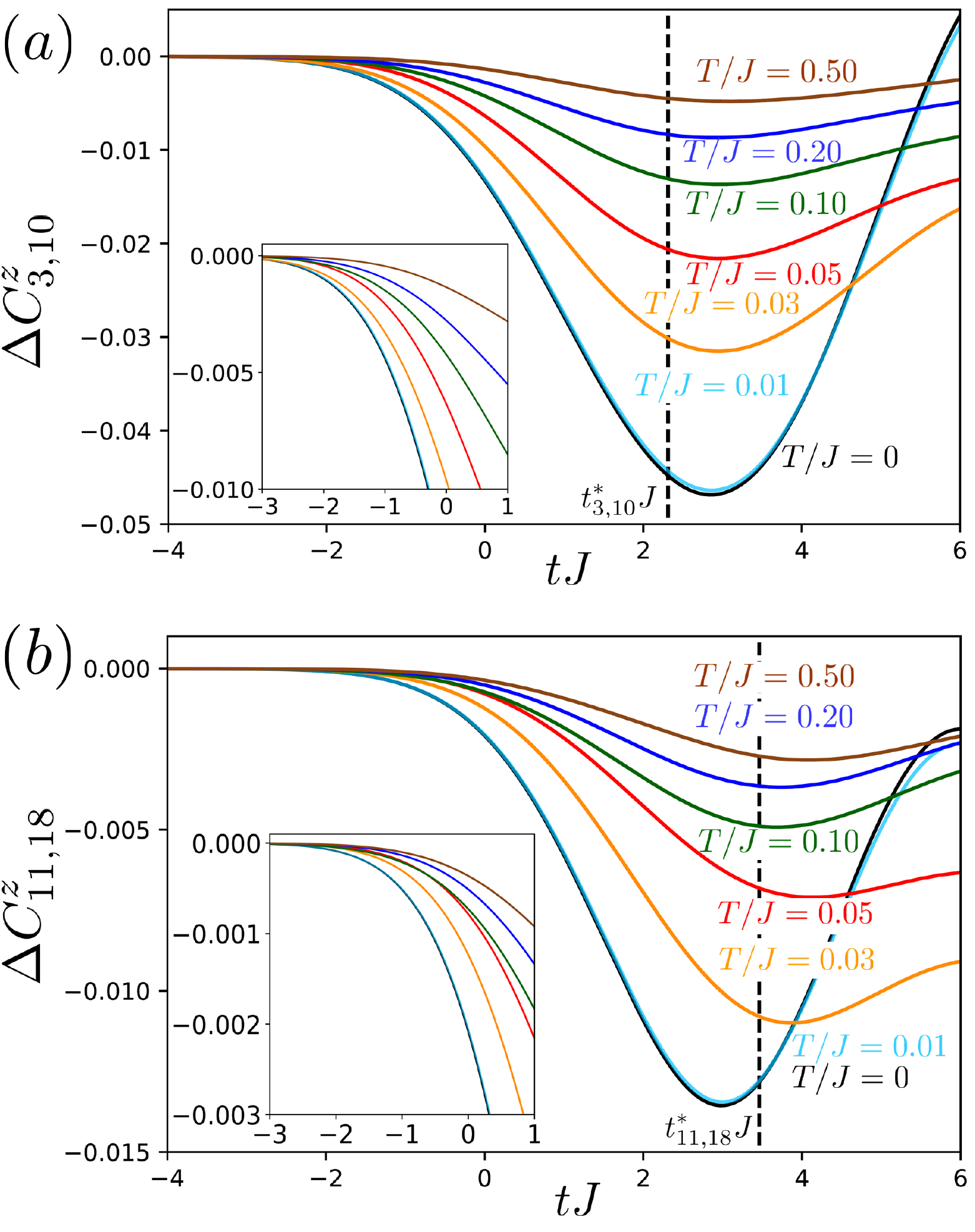}
  \caption{Real-time evolution of the change in the spin-spin correlation
    in the M region at several temperatures.
    The results at zero temperature are obtained
    by the exact diagonalization.
    Dashed vertical lines represent the time $t^*$ for the corresponding bonds.
    The insets of $(a)$ and $(b)$ are corresponding magnified graphs at $-3 < tJ < 1$
    with the same magnification.
  }
  \label{fig:bond}
\end{figure}

To clarify how the Majorana-mediated spin transport is modified at finite temperatures,
we first focus on the time evolution of the nearest-neighbor
spin-spin correlations $C_{ij}^\mu(t)$ on the $\mu$-bond.
This quantity is proportional to the bond energy, and thereby
the oscillation indicates the energy flow for the Majorana-mediated transport~\cite{KogaSpin_2020}.
Figure~\ref{fig:bond} shows the real-time evolution of the change
in the spin-spin correlations in the M region.
We find that at $T=0$, the oscillations of $\Delta C_{3,10}^z(t)$ and $\Delta C_{11,18}^z(t)$
start at $t \sim -2.5J$ and $t \sim -1.5J$, respectively  [see insets of Fig~\ref{fig:bond}].
This difference in time means that the energy injected by the magnetic pulse
in the L region in turn flows through the M region,
which is contrast to no oscillations in the magnetic moments [see Fig.~\ref{fig:lowT}].
For both bonds, the oscillation is little changed at $T/J<0.01$.
On the other hand, at $T/J\gtrsim 0.01$, it is rapidly changed, and
its intensity monotonically decreases with increasing temperatures.
This behavior seems a general feature in the correlated systems,
where the propagation smears due to thermal fluctuations.

\begin{figure}[t]
  \centering
  \includegraphics[width=\linewidth]{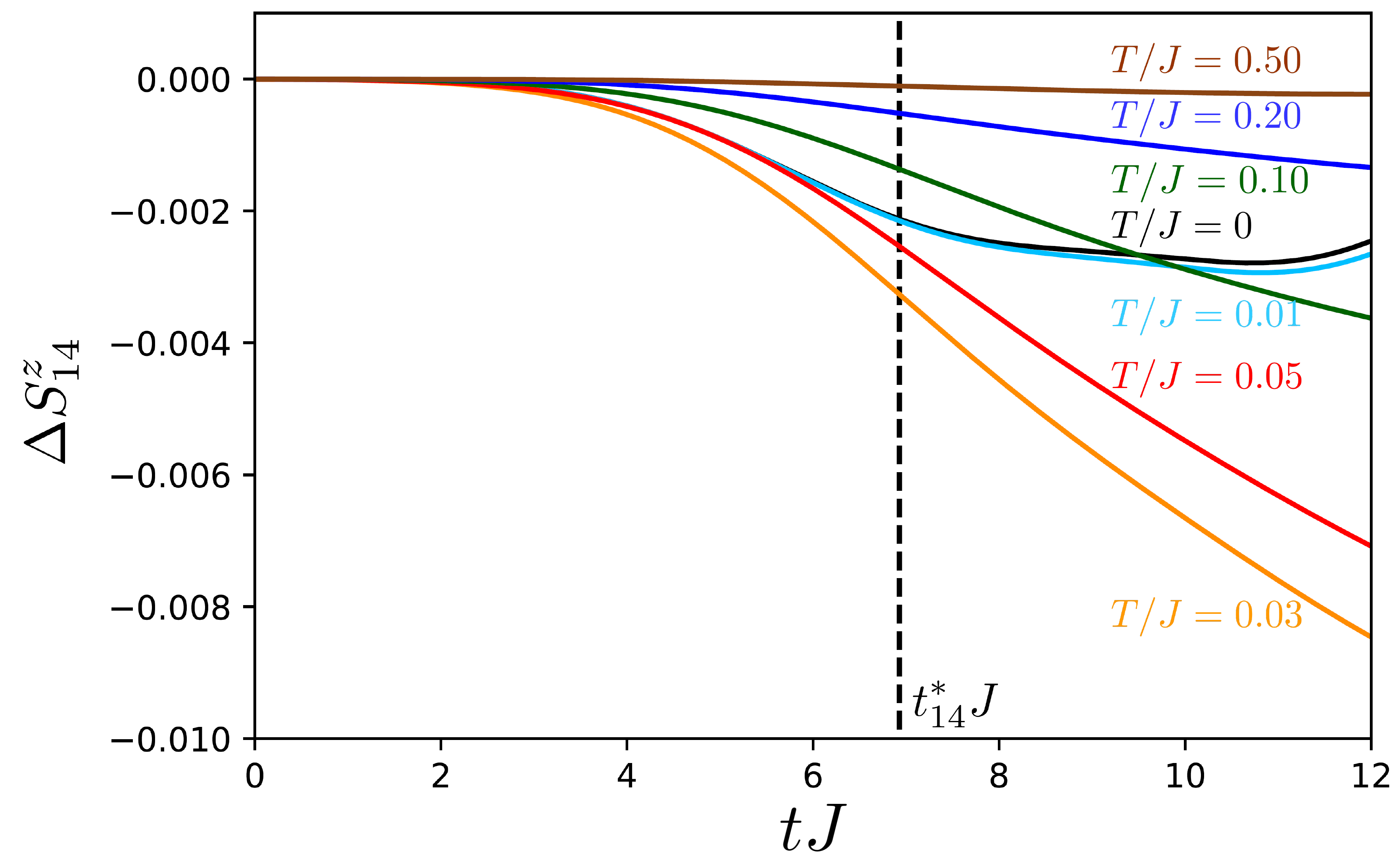}
  \caption{Real-time evolution of the change in the local magnetization $\Delta S^z_{14}(t)$
    in the R region.
    The dashed vertical line represents the time $t^*_{14}$.
  }
  \label{fig:highT}
\end{figure}

By contrast, different behavior appears in the change in the magnetic moment
in the R region.
The results with several temperatures are shown in Fig.~\ref{fig:highT}.
At low temperatures ($T/J \lesssim 0.01$), shoulder behavior in $\Delta S^z_{14}$ appears around $t=t^*$.
Beyond $T=T_L$, shoulder behavior smears and
the time evolution becomes monotonic.
The increase of $|\Delta S_{14}^z(t)|$ implies
that spin oscillations triggered by the magnetic pulse
are enhanced by thermal fluctuations.
With further increasing temperatures,
its magnitude decreases and almost vanishes
when $T\gtrsim T_H$.
This should originate from thermal fluctuations for both $Z_2$ fluxes and itinerant Majorana fermions.

\begin{figure}[t]
  \centering
  \includegraphics[width=\linewidth]{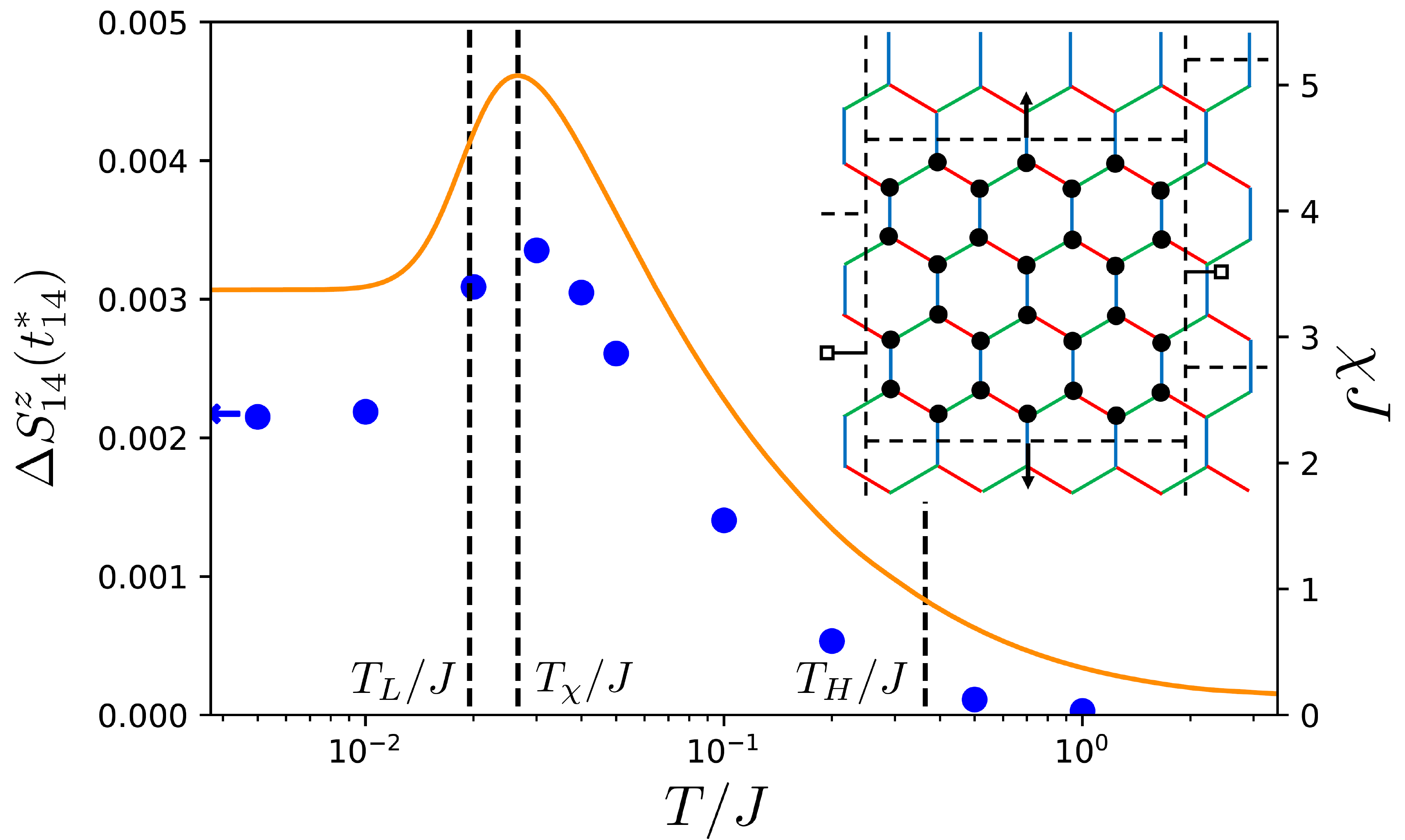}
  \caption{The blue circles represent $\Delta S^z_{14}(t)$ at $ t = t^*_{14}$
          and the orange line represents the magnetic susceptibility of the $N=28$ cluster (see the inset).
          The blue arrow represents the result in the ground state obtained by the exact diagonalization.
          Three dashed vertical lines indicate $T_L/J$, $T_\chi / J$, and $T_H/J$.
}
  \label{fig:chi}
\end{figure}

Now, we discuss the oscillation of the local moment in the R region in more detail,
regarding $|\Delta S_{14}^z(t)|$ at $t=t^*_{14}$ as its representative magnitude.
Figure~\ref{fig:chi} shows the temperature dependence of the quantity (blue circles).
It is found that, at zero temperature,
Majorana-mediated spin transport appears
with $|\Delta S_{14}^z(t^*_{14})|\sim 0.0022$.
This value little changes when $T/J\lesssim 0.01$.
With increasing temperatures, $|\Delta S_{14}^z(t^*_{14})|$ increases and takes a maximum around $T/J\sim 0.03$.
Further increase of the temperature decreases the value monotonically due to thermal fluctuations.
This nonmonotonic behavior reminds us of the magnetic susceptibility of the bulk system
~\cite{Singh_2010, Kubota_2015, Sears_2015, Freund2016, Yoshitake_2016, Yoshitake_2017}.

Then,
we consider a 28-site Kitaev cluster with the periodic boundary conditions along the $x$ and $y$ directions [see the inset of Fig.~\ref{fig:chi}].
Applying the TPQ method to this cluster with the tiny uniform magnetic field $h(=0.01J)$ in the $z$-direction,
we obtain the static susceptibility $\chi = \sum_i^N S_i^z / (hN)$.
We show the results with the solid line in Fig.~\ref{fig:chi}.
One can see that the susceptibility behaves non-monotonically against the temperature and shows a broad peak around $T=T_\chi(\sim 0.027J)$.
This non-monotonic behavior originates from two competing effects.
One is the thermal fluctuation, which tends to suppress the susceptibility.
The other is the existence of the finite gap of the $Z_2$ fluxes.
Due to the spin fractionalization, one spin is represented by the flux and itinerant Majorana fermion.

The flux becomes thermally activated at finite temperatures, in particular around $T_L$, which makes the spin
sensitive against the external field.
Our results show that the temperature dependence of the magnetic oscillations induced by the magnetic pulse and the susceptibility are essentially the same.
This implies that, also for the spin transport, the competition between the thermal fluctuations and the thermal activation of $Z_2$ fluxes plays a role similar to the case of the susceptibility,
leading to a non-monotonic temperature dependence.

\section{Conclusion}\label{sec:conclude}

We have investigated how stable the Majorana-mediated spin transport
in a quantum spin Kitaev model is against thermal fluctuations.
The finite-temperature spin dynamics has been examined
by applying the time-dependent thermal pure quantum state method to the Kitaev model.
At low temperatures $(T\ll T_L)$, an almost flux-free state is realized and
the spin excitation propagates similarly to that for ground state.
When $T\sim T_L$, larger oscillations in the spin moments are observed,
comparing with the results at the ground state.
At high temperature ($T \sim T_H$),
both itinerant Majorana fermions and localized $Z_2$ fluxes strongly fluctuate thermally, which suppresses the spin oscillations.
Our results demonstrate a crucial role of thermal fluctuations
in the Majorana-mediated spin transport.

We have found the enhancement of the spin oscillation around $T\sim T_\chi$
by considering the Kitaev model of a finite cluster.
In future, it is important to clarify how robust this non-monotonic behavior is against the system size
and whether it survives in more realistic setups with the Heisenberg terms  and/or disorders.

\acknowledgements
Parts of the numerical calculations are performed
in the supercomputing systems in ISSP, the University of Tokyo.
This work was supported by Grant-in-Aid for Scientific Research from
JSPS, KAKENHI Grant Nos.
JP19H05821, JP18K04678, JP17K05536 (A.K.) and JP20H05265, JP20K14412, JP21H05017 (Y.M.),
and JST CREST Grant No. JPMJCR1901 (Y. M.).

\bibliography{main}

\end{document}